# Relating land use and human intra-city mobility

Minjin Lee[1]

Petter Holme[1]

[1]Department of Energy Science, Sungkyunkwan University, 440-746 Suwon, Korea

E-mail adress: holme@skku.edu

## Abstract

Understanding human mobility patterns—how people move in their everyday lives—is an interdisciplinary research field. It is a question with roots back to the 19th century that has been dramatically revitalized with the recent increase in data availability. Models of human mobility often take the population distribution as a starting point. Another, sometimes more accurate, data source is land-use maps. In this paper, we discuss how the intra-city movement patterns, and consequently population distribution, can be predicted from such data sources. As a link between land use and mobility, we show that the purposes of people's trips are strongly correlated with the land use of the trip's origin and destination. We calibrate, validate and discuss our model using survey data.

## Introduction

Understanding human mobility patterns in urban areas is important in many fields—from city planning, via sociology and geography to complex systems science. In the last decade, there has been an increasing availability of human mobility data. This has enabled researchers to study mobility patterns in new ways [1]. Mobility is one of the aspects of human behavior that could be described as a law—albeit a statistical one [2]. This goal dates back to Ravenstein's 1885 paper [3] of a slightly different problem—how people move their home. Attempts to formulate a law, or model, of human mobility include the *gravity model* [4,5], the *intervening opportunities model* [6], *the rank-based model* [7], and the *radiation model* [8,9]. In practice, the above-mentioned mobility models focus on cities or countries that have a high mobile phone, GPS tracking or Internet use. These are, furthermore, usually politically stable, mature and developed regions. Several papers have pointed out this drawback [10–12]. Only few studies about human mobility have been done for developing areas [12,13]. This is a reason to develop methods to infer the mobility from other origins than population density. The advantage with land-use data as an input to mobility models is that it can to a great extent be inferred from satellite images (in practice complemented by national census data). Therefore, in principle, it should be rather easy to keep the same standard across different societies [14].

In this paper, we investigate how human mobility can be predicted, not from maps of the population distribution, but from land-use maps. To our knowledge, our work is the first with that goal. Older studies [15–17] have studied special cases of how land-use generates traffic, but not generalized these to a full mobility model. For example, Voorhees [15] study the "pull" certain land-use types have on the traffic along a road (but would need a models for how traffic is generated and how it branches at an intersection to make is a model for the same purpose as us).

Recent human mobility studies argue that we cannot hope for a law of mobility that holds accurately both for inter- and intra-city mobility—these phenomena seem to be fundamentally different. One observation (consistent over several data sets, including public transportation in Singapore [18], mobile phone data [10,19], and social media data [20]) is that intra-city human mobility has a high spatiotemporal regularity of the trajectories of individual. One reason for this is the daily routine of people [19–21]—commuting e.g. from their home, to work, to grocery shopping, etc. This is a phenomenon less frequent in inter-city travel (although it does exist, for example people returning home during holidays). Model of intra-city mobility can exploit the regularities of people's trajectories. Ref. [22], for example, derived an "activity-based" model that combines a Markovian transmission probability with a distance dependence to reproduce empirical mobility data.

As alluded to above, intra-city mobility is influenced by land-use patterns (which itself is a complex result of both the natural and built environment). In the urban planning literature, there are several studies relating land use and urban travel patterns. These studies mainly focus on the relationship between travel behaviors and summary statistics of the land-use patterns. The summary statistics measure properties such as how well mixed the spatial configuration of land-use types, or compactness a city's layout, is [23–27]. Some studies apply the relationship to build integrated land-use and transportation models like the commercial TRANUS (http://www.tranus.com/). This is a complex simulation model that divides the system into subsystems like activity demands, real estate supply, infrastructure supply and transport service demand to generate a state of transport equilibrium [28,29]. Other urban mobility studies try to elucidate the function of urban areas through mobility data [30,31]. Furthermore, Ref. [32] suggests a probability model to infer human activities using semantically enriched geographic data including land use data.

In this paper, we propose a model that takes land-use maps as input and predicts human mobility patters. It could be combined with population density maps, or other data sources, to increase its predictability. We begin our investigation by discussing the general relationship between land use and mobility. Second, we address how to predict mobility patterns using land-use maps. As a test case, we use the Origin-Destination (OD) survey data of the Greater Chicago region [33]. Finally, to validate our model, we match the population and trip-length distributions resulting from the predicted mobility patterns to the real population distribution of the Greater Chicago area.

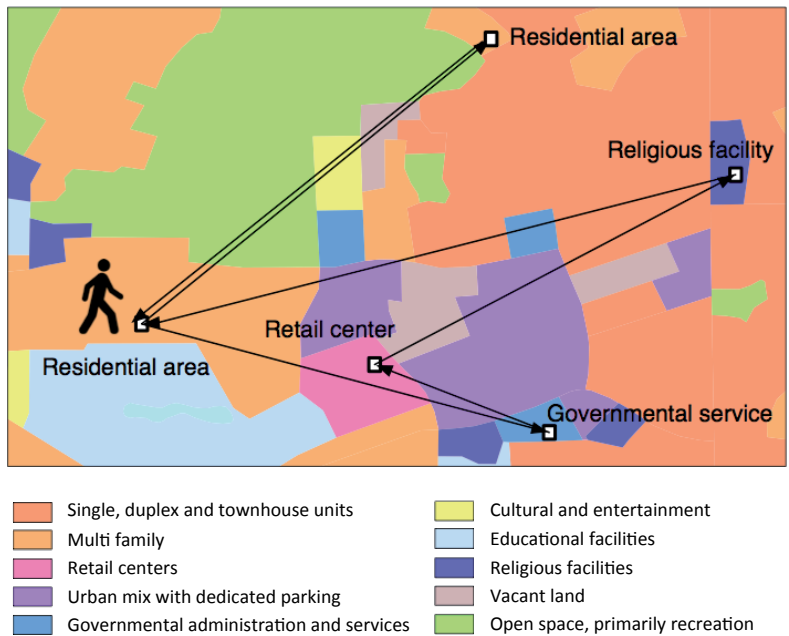

**Fig 1. Illustration of an individual's trajectory**

The example is taken from the Chicago origin-destination survey on a land-use map. It shows one person's one-day trajectory.

Figure 1 summarizes the main ideas of the method. We link individual trips with the land-use types at the origin and destination. The land-use type describes the primary role of an area with respect to human activity—if it contains entertainment facilities, retail centers, housing, or if it is a lake, forest, road, etc. It is thus a combination of information about the physical objects at the location and the activities people perform there. It should logically be—and we will illustrate that it is—connected to the purpose of people's trips (the *trip purpose*, as annotated in the OD data).

## Results

**Land use dependence of origins and destinations**

Our first analysis aims at relating land use and human activity. We use the greater Chicago region land-use map [34] and the OD data [33] of same region. As mentioned, in addition to the time, origin and destination, the OD survey data also records the purpose of the trips. It covers trips of 38,745 residents of the greater Chicago area for one or two days (some people participate in the survey for two days, others for one day), covering in total 180,200 trips. The mapping from the OD study to the land-use map is not entirely straightforward—technical details can be found in the *Methods* section. In Fig. 2, we show the fraction of trips to an area of a specific land use for each trip purpose.

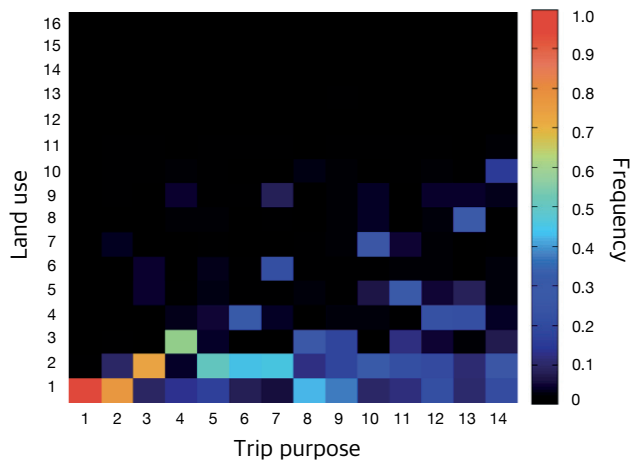

**Fig 2. The fraction of destination land-use type per trip purpose**

The trip purpose categories are as follows: 1. home activities, 2. visiting friends or relatives, 3. eating out, 4. attending class, 5. household errands or personal business, 6. service private vehicle, 7. shopping, 8. dropped off or picked up passenger, 9. civil and religious activities, 10. health care, 11. work related, 12. business related, 13. change of type of transportation, 14. recreation and entertainment. The land-use types are as follows: 1. residential, 2. urban mix, 3. educational, 4. government institutes, 5. office and business park, 6. shopping, 7. medical and health care facilities, 8. entertainment, 9. manufacturing and warehouse, 10. open space for leisure, 11. forest, 12. construction, 13. agricultural, 14. river, 15. transportation related, 16. social infrastructure. The land-use types are aggregated from 50 categories in the original data (how is explained in the *Methods* section).

Some activities are strongly connected with certain land-use types. For instance, trips with the purpose "home activities"—which is the largest class of trip purposes—are almost exclusively directed towards the land-use "residential". Other examples of strong correlations include that people traveling to "attend a class" tend to go to the land-use type "educational facilities" and people who traveling for "shopping" are likely to go to the land-use types "urban mix" or "shopping". Other types of trip purposes, such as "work related", "business related", "change type of transportation" or "recreation and entertainment" can go to a variety of land-use types. Furthermore, note that the number of trips to a certain land-use type is broadly distributed—land-use types "residential" and "urban mix" is the destination of 63% and 18% of the total number of trips, respectively, while the land-use types "forest", "construction", "agricultural", "river", "transportation" and "social infrastructure" all together covers 2% of the total number of trips. To put these numbers in context, we note that land-use type 1 and 2 covers 29% and 2% of the total area, respectively, while 11–16 covers 45%—the areas can explain some, but not all, of Fig. 2.

In conclusion, Fig. 2 tells us that the correlation between the land use and trip purpose varies from very strong to non-existent. However, most of the trips go to highly predictable land-use types (given their purpose), which means that overall there is much information in the land-use map that could be exploited to predict the trips, given their purpose.

**Land-use transition matrix**

Expanding the idea that land use reflects people's activity, we now turn to building a mobility model. We start by measuring how people travel between land-use types in the Chicago OD survey. We trace the trips on the land-use map (cf. Fig. 1) and count the pairs of land uses at the origin and destination. Then, we normalize these counts to obtain a transition matrix **L**. More precisely, the element $L_{ab}$ of **L** is defined as

$$L_{ab} = \frac{N_{ab}}{N} \frac{1}{A_a A_b} \tag{1}$$

where $N_{ab}$ is the number of trips originating at land-use type $a$ with a destination at land-use type $b$, $N$ is the number of all trips and $A_a$ is the total area of land use $a$. This means that our model assumes a first-order Markov process, ignoring the sequence of the trips. A higher-order model could include the land-use types of the origins and destinations of consecutive trips.

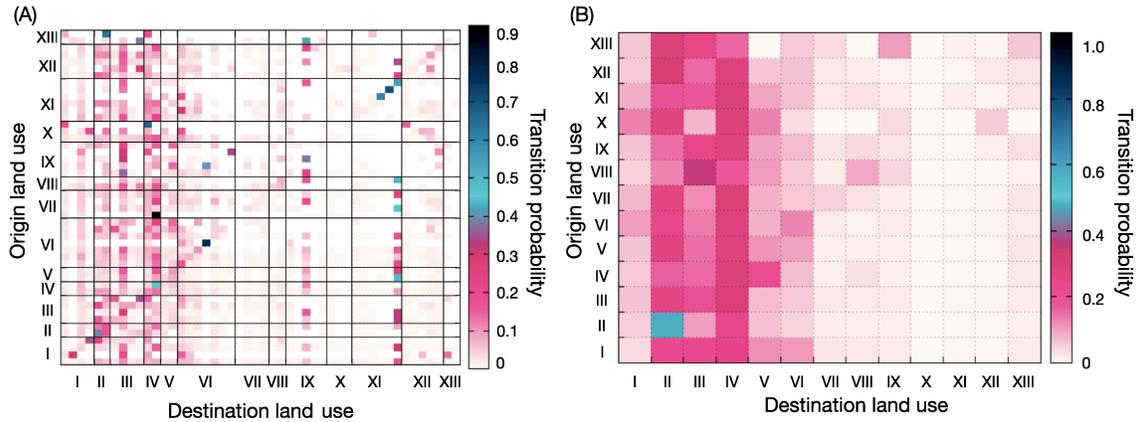

**Fig 3. The land-use transition probability matrix in two levels of aggregation**

(A) shows the original categorization, (B) shows our coarser categorization (note that it is a different aggregation than in Fig. 2 which is explained further in *Methods*). The land-use types in this aggregation are: I. residential, II. primarily retail and service, III. primarily office and professional, IV. urban mix, V. other commercial and service, VI. institutional, VII. industrial, VIII. transportation, IX. other transportation, communication and utilities, X. agricultural land, XI. open space, XII. vacant, wetlands or under construction, XIII. water.

In Fig. 3A, we show the transition matrix of the Chicago data as a heat map with the original categorization of land-use types (not the coarse grained version of Fig. 2). A conspicuous feature of Fig. 3A is that some destination land-use types are more frequent than others, regardless of the origin. This is manifested as vertical stripes of darker color. The horizontal marginal sums are one by construction. There are some examples of land-use categories that have a heightened probability of trips with the same origin and destination land-use type—for example "mineral extraction". There are two factors that can give a land-use type high values of the vertical marginal sum—either that it has many trips to it (the numerator is big) or the area (denominator) is small. Land-use types with a small area are more sensitive to fluctuations, meaning some $L_{ab}$ values would be lower if the survey had more samples. The large fluctuations for small area land-use types will not be a problem when we later use the transition matrix for our mobility model because these land-use types will be weighted down precisely because they have a small area.

Even though the fluctuations of **L** (Fig. 3A) is not a problem for the mobility calculations, it makes **L** itself harder to interpret. To get around this problem, we coarser categorization of the land-use types (also taken from the dataset itself). This coarse-grained transition matrix is displayed in Fig. 3B, where we can see that regardless of the origin land-use types, the "primary retail or service", "office or professional" and "urban mix" categories are the most popular destinations.

Note that the transition matrices are asymmetric, that is, the probability of a trip from *a* to *b* could differ from the probability of trips from *b* to *a*. This phenomenon reflects that people follow trajectories in their routines that are longer than just to a destination and back again (which is observed before in other data sources [21]).

**Mobility model**

Now we turn to a model of mobility patterns. Our model rests on four assumptions, all justified by observations. First, we assume that there are regularities in people's travel patterns. How regular the movements are and how they generalize across cultures affect the accuracy of the model. However, any regularity would contribute to a better prediction. Second, we assume that the land-use captures predominant human activities in an area, so that the land-use transition matrix reflects the regularities in travel patterns. Third, we assume that all locations (technically speaking, 700m × 700m squares of a square grid) of the map with the same land-use can be treated as equal. This assumption could of course be omitted if one has more detailed information (like additional estimated of the population density). Finally, we assume that trips between origin-destination pairs of the same combination of land uses follow the gravity model [4,5].

The gravity model states that the probability a person travels between two points separated by a distance $d$, is proportional to $1/d^\sigma$. The exponent $\sigma$ that can vary much between regions and type of transportation and has been measured from 0.3 to 3.0 [1]. We can write the gravity law as a relation for the expected flux between two locations $i$ and $j$.

$$T_{ij} = \frac{m_i m_j}{\left(d_{ij}/d_0\right)^\sigma} \qquad (2)$$

where $m_i$ is the attraction force (sometimes "mass" by analogy to physics) of the location $i$ and $d_0$ is a auxiliary parameter of the dimension length to make the denominator dimensionless (we set it to the linear dimension of a location, 700 m). It is common to use the residential population of a location as the attraction force. In our approach, we calculate it by a *land-use transition matrix* **L** (Fig. 3) that describes the trip flux between land-use types. In this way, we can, for example, account for the flux of students from the "residential" land-use type to the "educational", something purely population-density based models would cannot do. The trip flux from $i$ to $j$, $T_{ij}$, is determined from **L** and the power-law distance decay of the gravity model, as

$$T_{ij} = \frac{1}{A_j} \frac{t_i p_j}{\left(d_{ij}/d_0\right)^\sigma} \qquad (3)$$

where,

$$A_i = \sum_j \frac{L_{ij}}{\left(d_{ij}/d_0\right)^\sigma} \qquad (4)$$

and the number of trips from location $i$, $t_i$, is calculated by $t_i = T_{tot} p_i$, where $T_{tot}$ is total number of trips and $p_i$ is the *effective population* density at location $i$ (this quantity is also known as "average daily population" [35] or "strength" [36]). The effective population density is defined as the average fraction of the population present at a location during a day. In an urban area, the population within a location is constantly changing. For this reason, the residential population is not optimal for the purpose of calculating e.g. the trip-length distributions [35,36], which motivates the use of the effective population density. Note the recursive nature of the effective population density and the trip flux—the population density comes from the flux and the flux comes from the population density. Because of this recursive relationship, we do not try to replace the gravity model by the radiation model (which would otherwise have been interesting as it could potentially predict trip flux statistics better than the gravity model [8,9,37]). Since the radiation model needs to integrate the population around the origin and destination which would, at the present, be too time consuming for this study.

In order to calculate the $T_{ij}$, we need to estimate the effective population density distribution, $p_i$. For this purpose, we use the *location-to-location transition matrix* **M** that describes the trip flux between locations. **M**'s elements are given by

$$M_{ij} = \frac{L_{ij}}{(d_{ij}/d_0)^\sigma A_i} \quad (5)$$

Assuming that the population is continuously moving according to the Markov process defined by **M** gives the average population as the stationary distribution of the process. This is simply the solution to the equation $p\mathbf{M} = p$ (i.e. the eigenvector corresponding to the unity eigenvalue) [38]. Below, we calculate the $T_{ij}$, using the Eq. (3) for all pairs of cells. In that calculation, we assume the total number of trips out of a location to be proportional to the effective population density, i.e. will set $t_i$ to $p_i$.

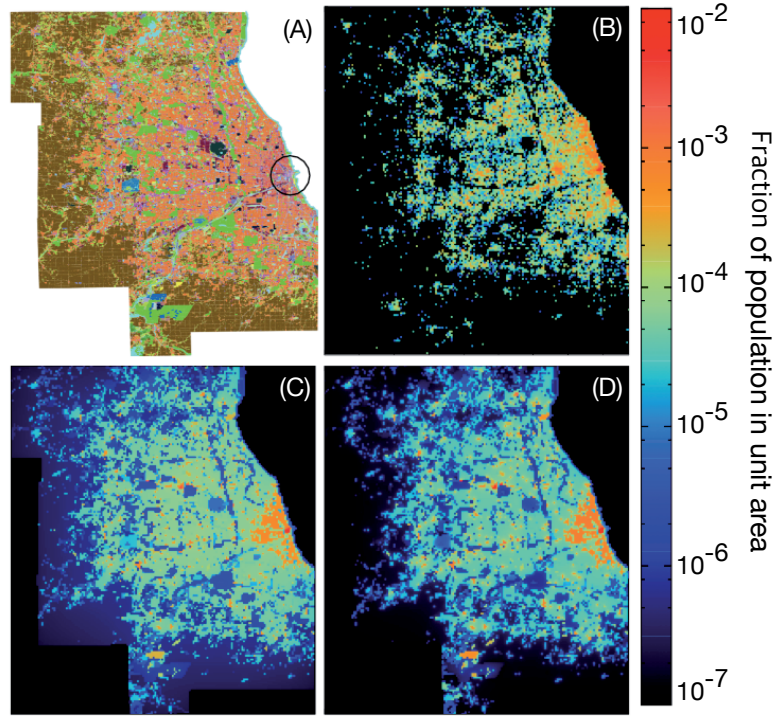

**Fig 4. Comparison of population density distribution of empirical data and our model**

Panel (A) shows the Chicago land-use map. (B) displays the effective population density distribution from the Chicago OD survey data. Panels (C) and (D) shows the effective population density from our simulations with distance exponents σ = 1.5 and 2.5 respectively (the optimal value is in between—σ = 1.90±0.04). The resolution of the map (i.e. the dimensions of a grid cell) is 700m × 700m.

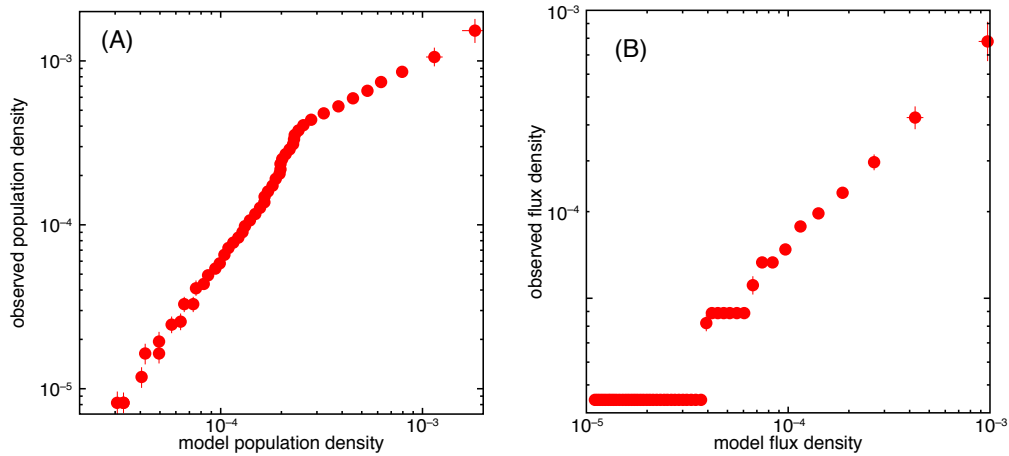

**Fig 5. Relation between empirical and predicted population density distribution and trip flux**

Panel (A) shows the average observed population density as a function of the population density predicted by the model. The data is binned into 50 segments such that each bin has an equal number of data points (although due to the log-scale the points with population density zero are not seen). Panel (B) show a similar plot (binned in the same way) for the trip flux computed according to Eq. (3). Note that both the quantities in (A) and (B) are non-dimensionalized by measuring the population in fraction of the total (i.e. dividing it by 25,845) and measuring distances in units of the grid cell side $d_0$ = 700 m. Error bars represent standard error. The step-like structure in (B) comes from location pairs with one, two, or three observed trips (the absence of error bars for these points is an artifact of the finite sample size).

Our first step to validate our mobility model is to compare the effective population density to the actual population density of the Chicago area (also calculated from the Chicago OD survey). These results are shown in Figs. 4 and 5. The simulation result captures many general features of the empirical population density distribution. The most crowded area (the circle area in Fig. 4A) is reproduced in both Fig. 4C and D and the outskirts are consistently sparse. Places where land-use types have lower attraction (rivers, forests, etc.) shows significantly lower population density in both empirical data and simulation results. Another observation is that as one increase the exponent values of the distance dependence, the population gets more centralized. Fig. 5A shows that the correlation between the predicted and measured population densities is not linear (although it is strong). We will not speculate in the reason of this nonlinearity beyond the mentioned approximations, the finite size of the system and the limitations of the gravity model. Fig. 5B confirms the correlations between the measured and observed trip fluxes.

The population distribution predicted by our model (Fig. 4C and D) and the empirical distribution we compare it to (Fig. 4B) are based on the same data source. However, the calculations are (as mentioned) very different and there is nothing *a priori* that says we would be able to recreate the same population density as accurately as we do, should our assumptions be wrong. We also notice that our results visually match other population data sources [40].

Finally, we notice that the simulation results do not completely match the empirical data. There are a number of reasons for that—the finite resolution of the land-use map, inaccuracies in that map, the higher (non-Markovian) effects of people's routines, a breakdown of the gravity model, etc.

**Trip length distribution**

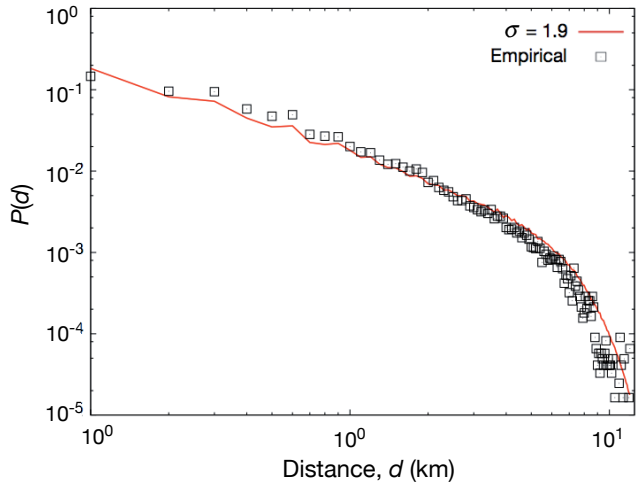

**Fig 6. Trip length distribution**

This figure shows the trip length distribution from the Chicago OD survey and a best fit to in by our activity model.

The trip length distribution from our simulation shows a strong similarity with the empirical data. See Fig. 6. Like other papers about trip length distributions for intra-city travel [39], our result also shows a power-law shape with an exponential cutoff. The cutoff (at around 80 km) comes from the spatial extent of the city. A gravity model with the population density as masses could also recreate this, but the population density is, evidently, not needed if one knows the land use.

As a goodness-of-fit criterion for the best exponent for the gravity model, we use the $\chi^2$ statistics (i.e. the sum of the squared difference between the empirical and the model data points)—see Fig. 7. As the two distributions are not independent, the assumptions for the chi-square test are not fulfilled and we cannot use $\chi^2$ for hypothesis testing. We can, on the other hand, use it to calibrate the model. Using a polynomial approximation close to the minimum of Fig. 7B, we estimate the minimum to 1.90±0.04. Liang *et al.* [35] use same data and the gravity model with the population density from a different source as attraction factor. They find the exponent 1.832 (no error estimate), which we take as a confirmation of our result.

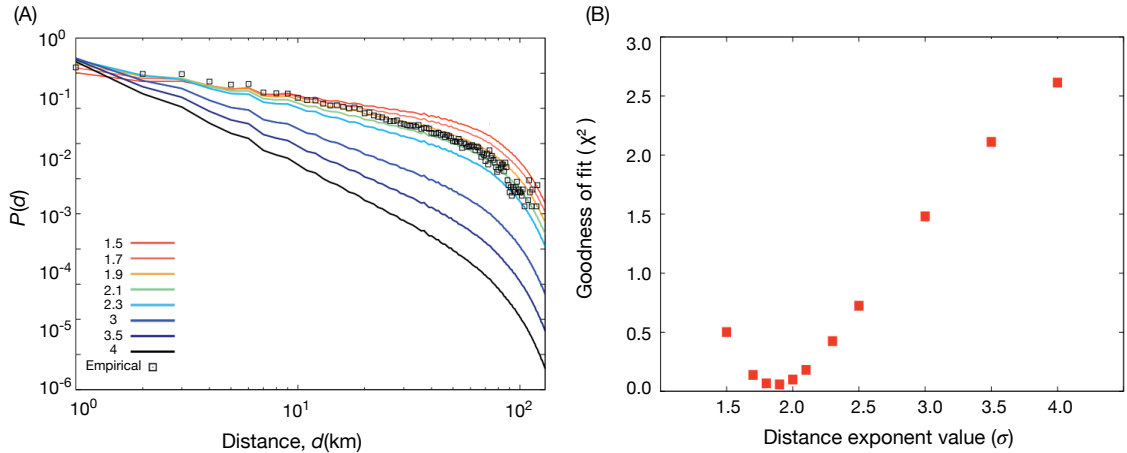

**Fig 7. Comparison of trip length distribution with different distance exponents**

In panel (A) we plot the trip length distributions with different distance exponents along with the empirical trip length distribution from the survey data. In panel (B), we show the values of the $\chi^2$ statistics as a function of the distance exponent value.

**Gravity model simulation with other models of population distribution**

Finally, we show that for simpler approaches, without the land-use map or without the empirical population distribution, our model cannot accurately predict the trip length distribution. We test the same model as outlined above but with an attraction factor coming from other population density estimates. We use the same exponent σ as obtained above. First, we replace the effective population density by an empirical population density. We have already concluded (in the context of Fig. 4), that these population densities are very similar, so (not so surprisingly) they give very similar trip-length distributions. Next we try a landscape of randomly assigned land uses, but with the same area distribution of the land-use types as in the real data—see Fig. 8. This population distribution creates, on average, longer trips. One way to understand this effect is that several land-use types have a tendency for trips to both start and end in it. At the same time there is a strong spatial auto-correlation in the map, manifested as compact areas of the same land-use type. Assume you are starting a trip with origin and destination in the same land-use type (i.e. the overrepresented situation), then you would have more places to choose from close by in the real landscape than in the random one. For the population distributions in Figs. 8C and D, on the other hand, the number of possible destinations is the same at all distances, and therefore we do not see the distance-increasing effect.

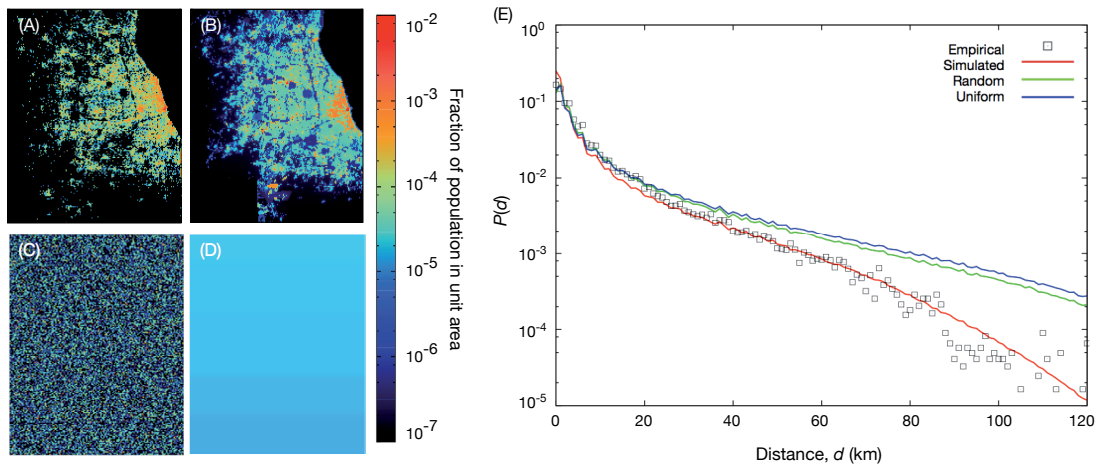

**Fig 8. Gravity model simulation with other population density models**

(A) shows results when we use the empirical population density distribution. (B) displays the effective population density calculated from the land-use maps. Panel (C) shows an instance of our random null model where the area per land use is conserved but everything else randomized. (D) illustrates a uniform population distribution. (E) is the trip length distribution of all four population distributions. For all curves, we use σ = 1.90.

## Discussion

We have investigated how intra-city travel is related to the land use of a city. First, we investigated the transition matrix between land-use types, capturing the trends. Then we combined these observations with the gravity model to a model of human mobility. We observe that this model not only predict the trip lengths but also the population density. We validate that the predicted population density, for the right choice of parameter of the gravity model, matches the observed population distribution accurately. The predicted trip length distribution can also closely reproduce the observed one, and this is not only an effect of the total areas of land-use types, but their spatial configuration.

A benefit of our model, is that the fluxes between locations can be (and, for our data, are) asymmetric. This is in contrast to the gravity model (and the radiation model too). It comes both from the asymmetry of the **L** matrix and the normalization factor $A_i$ (Eq. 3), and is observed in real data [35].

We have shown that there is as much information relevant for mobility prediction in the land-use map as in the population density. This leads to a simpler intra-city mobility prediction, especially cities where the population distribution is poorly mapped out. A caveat is that one needs to calibrate the model, specifically the land-use transition matrix, to empirical travel data (cf. Ref. [40]). This can, presumably be culture dependent, so using the same values for the transition matrix as we found could give a less accurate trip length distribution if applied to another city. It would be very interesting to explore this type of generalizability issues with other origin-destination surveys like the one we use (but we have not been able to obtain such at the moment).

Another caveat with our study is that we use the same data for calibrating and validating the model. The trip length distribution itself is a part of the model, the only thing it is used for it to find the transition matrix and the optimal exponent for the gravity model. In other words, the original 172,537 data points, which is the number of trips, is reduced to 50×50 transition matrix by the calibration, and then validated against full histograms. There is plenty of room for the prediction to go wrong, but it does work thanks to the land-use map having the necessary information. We show that a plain application of the gravity model cannot do as well as our mobility model. Another, consistency check is that the predicted population distribution matches other such estimates derived independently [41]. These issues focusing on the radiation model is discussed in Ref. [40].

Our method can be used to estimate the mobility with other geographical data, such as cellphone data or social media data. The characteristic spatial configuration can be obtained from, not only land-use map, but also location search services such as Yelp or Foursquare where there are both activity and spatial information. It would be interesting to infer mobility for cities in different countries by this type of data (which would be possible since those location services are used around the planet). Furthermore, there are several mobility studies using media data focusing on the activity patterns of people in urban areas [19–21]. Results from such studies, or from communication data [42,43], could also be combined with land-use maps to get a further understanding and better prediction of mobility patterns. Our model can also bring new insight to into studies focusing on urban planning for efficient land-use [28,29] and, if coupled to a model for urban planning, contribute to models of city growth [44,445].

In addition to the literature mentioned above, a longevous topic in the urban planning literature is how to control traffic flow and traffic congestion by planned land use changes [16,17]. Our model is directly applicable to simulation studies of this issue.

We believe there are many aspects of intra-city travel and urban structure left to be discovered. We have focused on trip lengths, but, for example, trip times and higher order properties of the trajectories of individuals could be interesting to explore.

## Methods

**Chicago origin-destination survey data**

The Chicago origin-destination survey data is downloaded from the CMAP (Chicago Metropolitan Agency for Planning) website [33]. This survey was conducted from January 2007 through February 2008. A total 10,552 households comprising 25,845 residents, living in northeastern Illinois, participated in either a one- or two-day survey. They provide a detailed travel inventory for each member of their household on the assigned travel day (or days). The survey covers demographics and travel behavior of the participants. This data set has four categories of data: personal data about the participants, (crude) location information of places that occur trajectories, the actual trajectories of the people, and the purpose of the trips (in 23 categories).

**Land-use map**

As mentioned, the land-use maps are sometimes closer to describing the activity people engage in at a location (like "entertainment") and sometimes closer to the physical composition of the land ("river"). Such a categorization is (perhaps inevitably) a bit arbitrary. For our purpose, the choice of categories is not crucial, but it probably plays an important role in how accurate the activity model is. The land-use map we use [34] covers northeast Illinois and contains 49 categories. It was published 2005 by the CMAP (Chicago Metropolitan Agency for Planning) and is downloadable at the website [34]. The land-use categories are assessed based on both visual inspection and software image processing. The format of the original map is an Environmental Systems Research Institute (ESRI) ArcGIS version 9.3 Geodatabase Polygon Feature Class. We converted this map into a 750m square-grid lattice where the land-use type of a grid cell gives the largest land use within the cell.

The original land use is classified into 9 main categories and 49 subcategories. For the land-use transition matrix, we use all 49 subcategories. However, to highlight the relationship between land-use categories and activity type in Fig. 2, we re-aggregate the categories. We maintain most of the main categories but make new aggregated categories for some frequently used land-use subcategories, e.g., shopping, office and business park, urban mix and entertainment which we assign to a new category called "commercial and service", and medical and healthcare facilities, educational, government institutes that we assign to a category called "institutional". We also aggregate categories for Fig. 3B, but for this figure we keep the original main categories.

**Mapping location to the land-use map**

We map the coordinates of locations onto the land-use map. The OD survey marks the locations with a centroid coordinate system, which is not precise enough for our purpose. Other studies have reported the same problem [32], but we found a way around it. The OD data typically also contains the name and street address of the origin and destination. We can use this information to achieve a higher precision coordinate via Google Maps' application programming interface [46]. To detect the land use corresponding to a coordinate, we use the ray-casting algorithm. It exploits ray-surface intersection tests to check whether a point is inside a given polygon [47].

## Acknowledgements

The authors thank Sang Hoon Lee for helpful comments and help with the location mapping. This research was supported by the Basic Science Research Program through the National Research Foundation of Korea (NRF) funded by the Ministry of Education (2013R1A1A2011947).